\documentclass{aastex61}

\usepackage{natbib}
\usepackage{url}
\usepackage{xspace}
\usepackage{booktabs}

\newcommand{\kk}{{\it{K2}}\xspace}
\newcommand{\kepler}{{\it{Kepler}}\xspace}
\newcommand{\tess}{{\it{TESS}}\xspace}
\newcommand{\dssi}{{\it{DSSI}}\xspace}
\newcommand{\nessi}{{\it{NESSI}}\xspace}
\newcommand{\alopeke}{{\it{`Alopeke}}\xspace}
\newcommand{\zorro}{{\it{Zorro}}\xspace}

\received{}
\revised{}
\accepted{}
\submitjournal{}

\shorttitle{Speckle Imaging \tess data}
\shortauthors{Matson et al.}

\begin{document}

\title{Detecting Unresolved Binaries in \tess Data with Speckle Imaging}

\correspondingauthor{Rachel Matson}
\email{rachel.a.matson@nasa.gov}

\author[0000-0002-0786-7307]{Rachel A. Matson}
\affil{NASA Ames Research Center, 
Moffett Field, CA 94035, USA}

\author{Steve B. Howell}
\affiliation{NASA Ames Research Center, 
Moffett Field, CA 94035, USA}

\author{David Ciardi}
\affiliation{Caltech/IPAC - NASA Exoplanet Science Institute
1200 E. California Ave,
Pasadena, CA 91125, USA}



\begin{abstract}
The {\it{Transiting Exoplanet Survey Satellite (TESS)}} is conducting a two-year wide-field survey searching for transiting exoplanets around nearby bright stars that will be ideal for follow-up characterization. To facilitate studies of planet compositions and atmospheric properties, accurate and precise planetary radii need to be derived from the transit light curves. Since $40-50$\% of exoplanet host stars are in multiple star systems, however, the observed transit depth may be diluted by the flux of a companion star, causing the radius of the planet to be underestimated. High angular resolution imaging can detect companion stars that are not resolved in the \tess Input Catalog, or by seeing-limited photometry, to validate exoplanet candidates and derive accurate planetary radii. We examine the population of stellar companions that will be detectable around \tess planet candidate host stars, and those that will remain undetected, by applying the detection limits of speckle imaging to the simulated host star populations of \citet{Sullivan:apj2015a} and \citet{Barclay:ArXiv2018a}. By detecting companions with contrasts of $\Delta m \lesssim 7 - 9$ and separations of $\sim0.02 - 1\farcs2$, speckle imaging can detect companion stars as faint as early M stars around A$-$F stars and stars as faint as mid-M around G$-$M stars, as well as up to 99\% of the expected binary star distribution for systems located within a few hundred parsecs.


\end{abstract}


\section{Introduction} \label{sec:intro}

Building on the success of NASA's \kepler Mission, the {\it{Transiting Exoplanet Survey Satellite}} (\tess) is an all-sky survey designed to find close-in transiting planets orbiting the nearest stars, with an emphasis on detecting small planets orbiting bright stars \citep{Ricker:JATIS2015a}. The planets discovered by \tess will be well suited for follow-up observations and atmospheric characterization via ground-based facilities and missions such as the {\it{Hubble Space Telescope}} and {\it{James Webb Space Telescope}} ({\it{JWST}}). However, the thousands of planet candidates detected by \tess will require substantial follow-up efforts in order to confirm their planetary status and identify the planets most suitable for atmospheric characterization \citep{Kempton:ArXiv2018a}. 

Planets discovered with photometric surveys provide a wealth of information via their light curves but require additional confirmation as other astrophysical phenomena can mimic the signal of a transiting planet. For instance, eclipsing binaries with grazing eclipses or whose light is blended with a candidate host star can have eclipses that resemble planetary transits \citep{Fressin:apj2013a}. Follow-up observations are therefore necessary to confirm or validate transiting planets, as false positives typically outnumber true planetary systems \citep{Torres:apj2011a}. High-resolution spectroscopy, in particular, can rule out close binary companions and measure the mass of a planet. However, for faint targets ($V \gtrsim 12$) and the sheer number of planetary candidates found by space missions like \kepler and \tess, spectroscopic confirmation of every planet becomes impossible. Alternatively, high-resolution imaging is used to detect additional stars within the photometric aperture that may be responsible for the observed transit or dilute a stellar eclipse so that it mimics a planetary transit \citep{Everett:aj2015a}. Statistical validation techniques are then used to asses the relative probabilities of planetary and false positive scenarios for a planet candidate, determining whether it is more likely to be a transiting planet or a false positive \citep[e.g.][]{Torres:apj2011a, Morton:apj2012a}. High-resolution imaging therefore rules out various false positive scenarios and enables robust validation of planet candidates. Such follow-up will be especially important for \tess as the $\sim$$21''$ pixels will often include multiple stars, increasing the risk of blends and false positives. While \tess is targeting nearby, bright stars and many of the detected exoplanets will be amenable to mass measurements through precision radial velocities, detecting contaminating stars will be vital for confirming the planetary nature of transit events and validating the thousands of planet candidates, including low mass and more distant planets.
 
Once a planet candidate is confirmed or validated, if the host star is single and isolated, the radius of the planet can be obtained from a simple fit to the transit light curve. Accurate and precise planet radii are invaluable for determining planetary distributions as well as addressing fundamental questions concerning the bulk compositions, atmospheric properties, and formation histories of exoplanets  \citep{Madhusudhan:prplVI2014a}. Planet mass and radius determinations allow for calculations of the bulk planet density, which constrains the interior composition of the planet. Incident flux, and therefore planetary equilibrium temperatures, also depend on planetary radii \citep{Johnson:aj2017a} and models of transiting Earth-sized planets have shown that the bulk composition of a planet strongly affects the resulting thermal emission and transmission spectra \citep{Morley:apj2017a}. Thus, in order to interpret the atmospheric signatures observed via transit and emission spectroscopy, accurate knowledge of the planetary radii are required. 

If a host star is not isolated, however, the observed transit depth will be diluted by the flux of the companion star, causing the radius of the planet to be underestimated \citep{Ciardi:apj2015a}. Both bound and line-of-sight companion stars can contaminate the light curve and affect the measured planetary radii. \citet{Furlan:aj2017a} found that $\sim$\,$30\%$ of \kepler Objects of Interest (KOIs) observed with high-resolution imaging had at least one companion detected within $4''$ (pixel scale of \kepler). In reality the true number of KOI companion stars is larger due to the varying sensitivity limits of the imaging techniques. Several studies of stellar multiplicity in \kepler and \kk exoplanet host stars have found companion fractions of $40-50\%$ \citep[e.g.][]{Horch:apj2014a, Deacon:mnras2016a, Matson:aj2018a, Ziegler:aj2018b}, consistent with solar-type stars in the solar neighborhood \citep{Raghavan:apjs2010a}. While other studies find fewer close binary companions around \kepler exoplanet host stars \citep{Wang:apj2014a, Wang:apj2014b, Kraus:aj2016a}, planet hosting stars, in general, appear to follow similar binary trends as field stars. We therefore expect comparable fractions of unresolved binaries and blended companions in \tess data, with the larger pixel scale potentially increasing the number of blended companions. Without properly accounting for the presence of stellar companions within the photometric aperture of \tess, planet sizes will be underestimated, potentially affecting both the density and surface gravity derived for the planet as well as the interpretation of atmospheric signatures. 

Such unresolved binaries also give rise to systematic errors in star and planet counts, which lead to biases in planet occurrence rates. In particular, the occurrence rates of planets with radii less than $\sim$$2 R_{\Earth}$ may be underestimated by as much as $50\%$ \citep{Bouma:aj2018a}, a key regime for understanding the formation and evolution of rocky super-Earths and sub-Neptunes with thick atmospheric envelopes \citep{Fulton:aj2017a, Owen:apj2017a}. High-resolution imaging of \tess exoplanet candidates therefore plays a vital role in detecting stellar companions and correcting planetary radii, necessary for deriving unbiased planet occurrence rates, bulk compositions of planets, and performing atmospheric characterization.

Seeing-limited images from ground-based, wide-field surveys, such as 2MASS or SDSS, can reveal stellar companions at near-equal contrast ratios within $\sim3''$ of exoplanet host stars \citep{Ziegler:aj2017a}, making them useful for identifying blended stars within the \tess pixels and ruling out nearby eclipsing binaries as the source of any transits. However, higher resolution imaging using adaptive optics, speckle, or lucky imaging is required to identify close companions and provide more precise confirmation and characterization of planetary systems. Speckle imaging, in particular, can detect companions around exoplanet host stars within $\lesssim0.02-1\farcs2$ and up to $\sim$10 magnitudes fainter than the host star \citep{Howell:aj2011a, Horch:aj2012a}. Detecting companions in this region is especially important as studies of \kepler stars have shown that most companions within $1''$ are bound, while only $\sim$50\% of companions at $2''$ are bound \citep{Horch:apj2014a, Hirsch:aj2017a, Matson:aj2018a, Ziegler:aj2018b}. As bound companions are closer to the host star, and tend to be roughly equal in brightness, 
they are more likely to dilute exoplanet transits and affect planetary radii \citep{Furlan:aj2017b}. In addition, nearly all \tess targets will be brighter and closer than the stars observed by \kepler, increasing the effectiveness of high-resolution imaging. Instead of probing within $\sim$100\,AU of the host stars, high-resolution imaging will be able to detect companion stars within $\sim$$1-10$\,AU of the host stars, thereby decreasing the fraction of undetected companions \citep{Ciardi:apj2015a}.

The purpose of this paper is to illustrate the importance of high-resolution follow-up imaging for \tess planet candidate host stars, which are nearly ten times closer than \kepler host stars, by identifying the parameter space in which speckle imaging can find unresolved stellar companions to aid in planet validation and characterization efforts. We begin by briefly discussing the capabilities of our speckle instruments and our follow up efforts for \kepler and \kk in Section \ref{sec:speckle}. In Section \ref{sec:tesssim}, we use the detection limits of speckle imaging to identify possible companions that can be detected around exoplanet host stars from the \tess planet yield simulations of \cite{Sullivan:apj2015a} and \cite{Barclay:ArXiv2018a}. Finally, the fraction of stellar companions we expect to detect for \tess exoplanet hosts as well as the role of speckle imaging in \tess follow-up efforts are discussed in Section \ref{sec:disc}.

\section{Speckle Imaging of Exoplanet \\Host Stars} \label{sec:speckle}

Speckle imaging removes the effects of turbulence in the atmosphere in order to reach the theoretical diffraction limit for a telescope with a given diameter and wavelength. Since 2010, the Differential Speckle Survey Instrument \citep[\dssi;][]{Horch:aj2009a} has been used to provide diffraction limited imaging of targets of interest from the \kepler/\kk mission at the 3.5\,m WIYN\footnote{The WIYN Observatory is a joint facility of the University of Wisconsin-Madison, Indiana University, the National Optical Astronomy Observatory, and the University of Missouri} telescope at Kitt Peak National Observatory, the 8\,m Frederick C. Gillett Gemini North telescope on the summit of MaunaKea in Hawaii, and the 8\,m Gemini South telescope on the summit of Cerro Pachon in Chile. Based on the success of \dssi, two new speckle instruments were designed and built at NASA's Ames Research Center to enable validation and characterization of exoplanet candidates discovered by \kepler, \kk, and future missions. The NASA-NSF Exoplanet Observational Research (NN-EXPLORE) Exoplanet Stellar Imager ({\it{NESSI}}) is available at WIYN \citep{Scott:pasp2018a}, while the second speckle instrument, {\it{`Alopeke}}, resides at Gemini North (Scott et al., in preparation). \dssi is still used as a visiting instrument at Gemini South, but a third dedicated speckle instrument, \zorro, will be commissioned there in summer 2019. All speckle instruments are available to the public, with reconstructed images and contrast limit curves produced by the speckle team and provided to the PI and/or the Exoplanet Follow-up Observing Program (ExoFOP) archives\footnote{\url{https://exofop.ipac.caltech.edu/}} following each observing run.

Each speckle instrument offers simultaneous two-color diffraction-limited imaging in the optical, using two EMCCD's and combinations of narrow-band ($40 - 50$\,nm wide) filters, to provide simultaneous photometric and astrometric data at subarcsecond precisions. This enables the identification of background objects and companion stars that contaminate exoplanet transit detections within $\sim 0.02 - 1\farcs2$ and up to $\sim$10 magnitudes fainter than the exoplanet host star. For any detected companions, speckle imaging provides the position angle, separation, and contrast from the host star, as well as color information that helps reduce the parameter space of false positives and correct exoplanet radii derived from blended binary sources \citep{Ciardi:apj2015a}. While \nessi and \alopeke are nearly identical instruments (containing the same filters and optical components; and similar to \dssi), the aperture size of Gemini is more than twice that of WIYN, resulting in greater light gathering power and a smaller diffraction limit. Table \ref{tab:spkcomp} illustrates the different detection limits in terms of $V$- magnitudes and magnitude contrast ($\Delta m$) at WIYN and Gemini, as well as the angular resolutions obtainable using the \nessi/\alopeke
narrow band speckle filters. More details of the design parameters and characteristics of \nessi and \alopeke can be found in \cite[][and Scott et al., in preparation]{Scott:pasp2018a} and on the instrument webpages of WIYN\footnote{\url{http://www.wiyn.org/Instruments/wiynnessi.html}} and Gemini\footnote{\url{https://www.gemini.edu/sciops/instruments/alopeke/}}. 

\begin{deluxetable}{lCC}[h]
\tablecaption{Speckle Imaging Capabilities of \nessi(WIYN) and \alopeke/\zorro (Gemini) \label{tab:spkcomp}}
\tablecolumns{3}
\tablenum{1}
\tablewidth{0pt}
\tablehead{
\colhead{} &
\colhead{WIYN} &
\colhead{Gemini}
}
\startdata
Typical magnitude limit ($V$) & 14 & 17 \\
Typical contrast limit ($\Delta m$)  & 6.5 & 7 - 9 \\
Diffraction limit at 467nm  & 0.034'' & 0.015'' \\
Diffraction limit at 562nm  & 0.040'' & 0.017'' \\
Diffraction limit at 716nm  & 0.051'' & 0.022'' \\
Diffraction limit at 832nm  & 0.060'' & 0.026'' \\
\enddata 
\end{deluxetable}

\begin{figure}[h]
\includegraphics[scale=0.65]{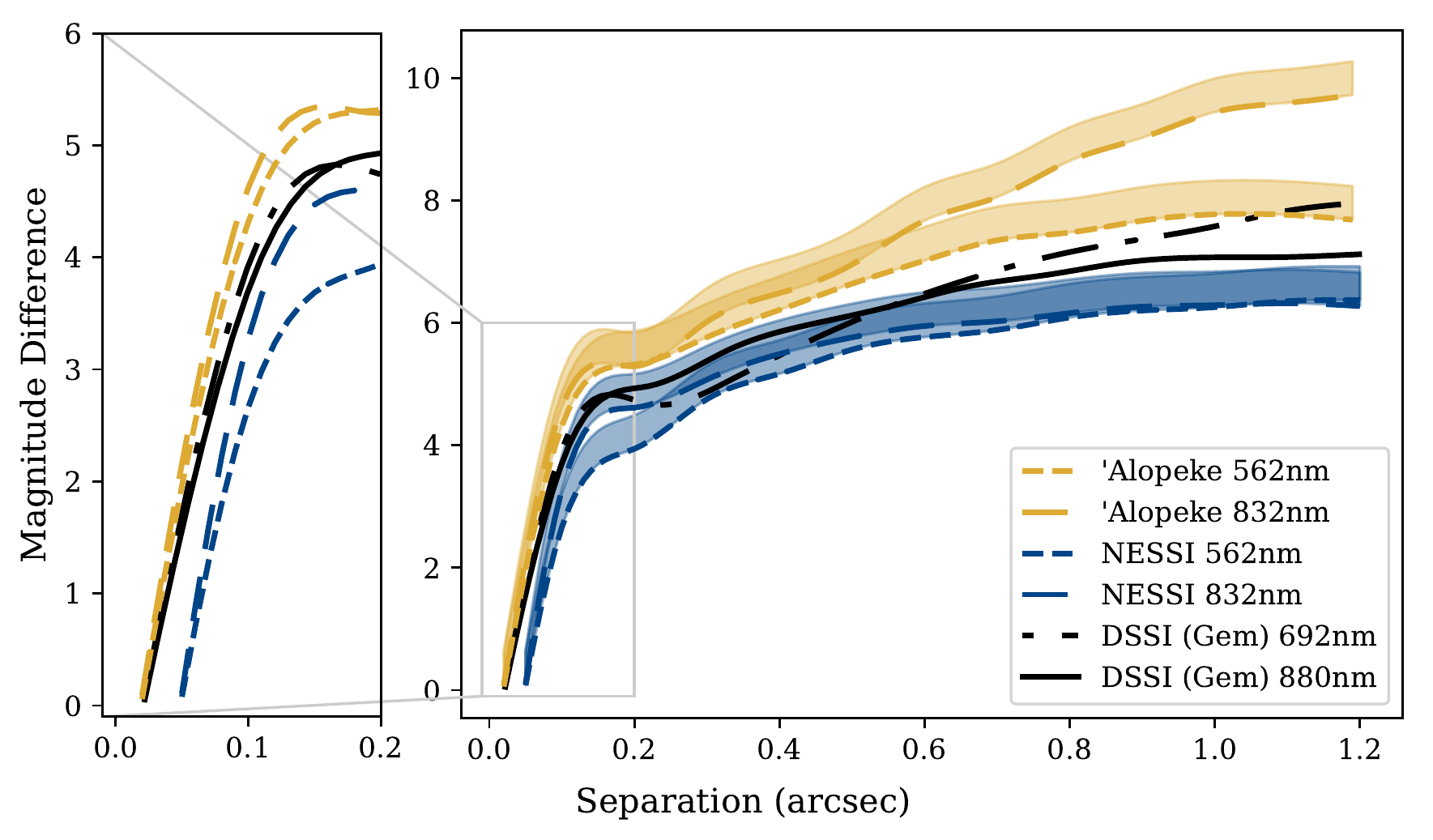}
\centering
\caption{Example $5\sigma$ detection limit curves for \alopeke at 562nm (dashed yellow line) and 832nm (long dashed yellow line), and \nessi at 562nm (dashed blue line) and 832nm (long dashed blue line), with shaded regions showing the $3\sigma$ detection limits. Also plotted are example \dssi detection limit curves from Gemini at 692nm (dot dashed black line) and 880nm (solid black line). The inset on the left highlights the difference in resolution achievable at WIYN (\nessi) and Gemini (\alopeke), and the rapid increase in sensitivity between $0-0.2''$ for speckle imaging. Stellar companions with a separation and delta magnitude that fall below a given curve are detectable with speckle imaging.
\label{fig:compcurves}}
\end{figure}

For high-resolution imaging it is crucial to determine the sensitivity of the images in terms of contrast limits and how the contrast depends on separation from the primary star \citep{Lillo-Box:aap2014a}. The sensitivity of speckle imaging to companions rises sharply from the diffraction limit to a `knee' at a separation of $0.15-0.2$ arcseconds, where it then continues to slowly increase out to $\sim$$1\farcs2$, beyond which the speckle patterns become de-correlated \citep{Horch:aj2012b, Horch:aj2017a}. While detection limits vary based on observing conditions and signal-to-noise ratios, on average speckle is sensitive to contrasts of $\sim5$~magnitudes at separations of 0.2~arcsec and $\sim6$~magnitudes at 1.0~arcsec at WIYN \citep{Horch:aj2017a}, and approximately 5 magnitudes at 0.1 arcsec separations and 7 - 9 magnitudes at 1.0 arcsec at Gemini \citep{Horch:aj2012a}. Example $5\sigma$ detection limit curves showing the magnitude contrast ($\Delta m$) as a function of separation for \alopeke at 562nm (dashed yellow line) and 832nm (long dashed yellow line), and \nessi at 562nm (dashed blue line) and 832nm (long dashed blue line), are shown in Figure \ref{fig:compcurves}. Shaded regions show the $3\sigma$ detection limits. Stellar companions with a separation and delta magnitude that fall below a given curve are detectable with speckle imaging. \dssi has slightly lower sensitivity than \alopeke, as shown by the Gemini detection limit curves at 692nm (long dashed black line) and 880nm (solid black line). The left panel of Figure~\ref{fig:compcurves} shows the $5\sigma$ detection limit curves zoomed in on the region $0-0.2''$ from the host star, highlighting the small angular separations detectable with speckle imaging and the rapid rise in sensitivity to fainter companions as a function of separation. In general, speckle imaging has higher angular resolution than near-infrared adaptive optics (AO) imaging and is therefore more sensitive to close companions, but has slightly lower sensitivity than AO to faint companions beyond $\sim$$0.2''$. In this regard, AO and speckle imaging are complementary techniques since most AO observations have relatively shallow detection limits within 0\farcs2, as they are made at longer wavelengths, while speckle imaging provides diffraction-limited imaging over a very small field of view at shorter wavelengths.
For binaries observed with \dssi at WIYN, the accuracy and precision of astrometry from speckle imaging has a typical precision less than 0.5$\degr$ in position angle and $1-3$ mas in separation \citep{Horch:aj2011b}, with analogous results obtained for \kepler targets observed at Gemini \citep{Horch:aj2012a}. 

Speckle observations of exoplanet host stars made significant contributions to the imaging portion of the \kepler Follow-up Observation Program \citep{Furlan:aj2017a} and have contributed directly to over 100 referred publications, having imaged over 80\% of the \kepler host stars and nearly 50\% of the currently known \kk candidates. These observations, publicly available at ExoFOP\footnote{{\url{https://exofop.ipac.caltech.edu/}}}, have produced thousands of host star characterizations, including the first Earth-size planet detected in a stellar habitable zone \citep[Kepler-186f;][]{Quintana:Science2014a} and four small, likely rocky planets that are candidates for transit spectroscopy with JWST (K2-3b, K2-9b, K2-18b, \citealt{Crossfield:apjs2016a}; TRAPPIST-1, \citealt{Howell:apjl2016a}). Observations of \tess planet candidates will extend such work to planets around a wide range of stellar types and orbital distances, while validating systems and preparing for detailed follow-up.

\section{Unresolved Stellar Companions in \tess}\label{sec:tesssim}

Several papers have simulated the expected exoplanet yield for \tess in order to inform target prioritization and plan for follow-up studies. \cite{Sullivan:apj2015a} simulated the population of exoplanets and eclipsing binaries that \tess will detect using the TRILEGAL galaxy model \citep{Girardi:aap2005a} to generate a catalog of stars and added planets according to occurrence rate statistics derived from the \kepler mission \citep[e.g.][]{Fressin:apj2013a,Dressing:apj2015a}. Since then, others have reexamined the expected \tess planet yield, including \cite{Bouma:ArXiv2017a} who explored extended mission strategies, and \cite{Ballard:ArXiv2018a} who created updated planet yields for M-dwarfs. In addition, \cite{Barclay:ArXiv2018a} reported revised estimates of the numbers and characteristics of exoplanets the \tess mission will find based on physical stars in the \tess Candidate Target List \citep[CTL;][]{Stassun:ArXiv2017a} and the latest \tess hardware specifications. 
In this work we use the \tess predictions of \cite{Sullivan:apj2015a} and \cite{Barclay:ArXiv2018a} to examine the expected host star properties and determine the types of bound stellar companions that will be unresolved in \tess data and their impact on planetary properties. Since we anticipate \zorro, which will be identical to \alopeke, to be available in early 2019 we focus on the observing capabilities of \alopeke/\zorro and \nessi; however, \dssi will make similar contributions at Gemini South in the meantime.

\subsection{Sullivan et al.~2015} \label{subsec:sullivan}

\begin{figure*}[ht]
\includegraphics[scale=0.6]{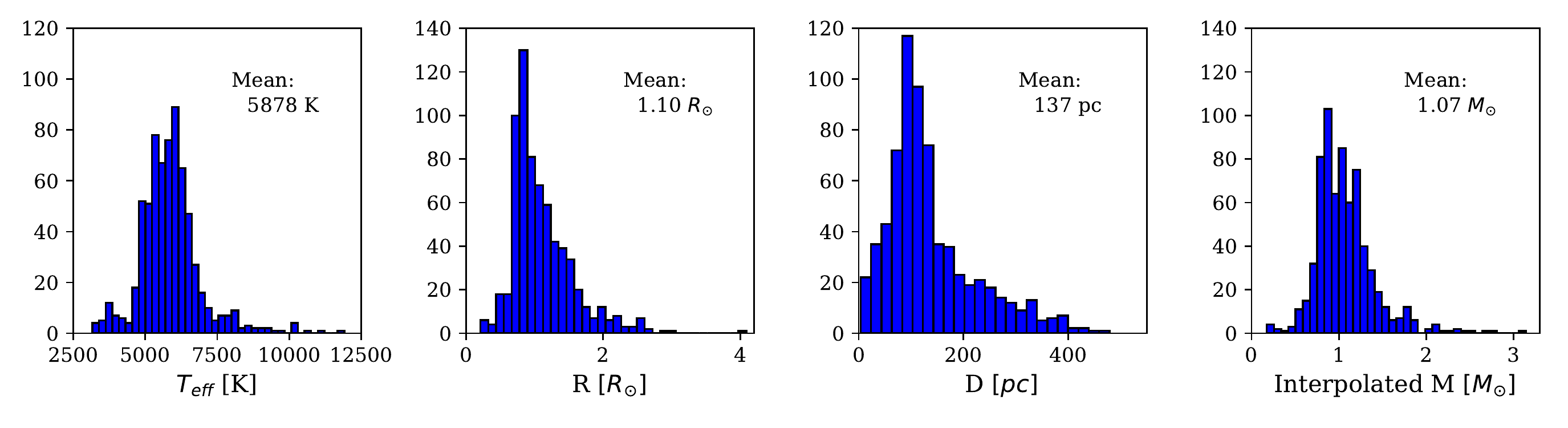}
\centering
\caption{Stellar properties of 682 simulated bright dwarf stars ($V \leq 12$) with exoplanets detectable by \tess from \cite{Sullivan:apj2015a}. The stellar masses are interpolated from the effective temperatures and radii given in \citeauthor{Sullivan:apj2015a} and the Modern Mean Dwarf Stellar Color and Effective Temperature Sequence Table based on \cite{Pecaut:apjs2013a}.
\label{fig:Shist}}
\end{figure*}

The stellar population simulated by \cite{Sullivan:apj2015a, Sullivan:apj2017a} includes $1.58 \times 10^{8}$ stars brighter than $K_s = 15$, $1.81 \times 10^{9}$ stars between $K_s = 15$ and $T = 21$ for which \tess could detect a deep eclipse from a binary star (used as background stars to create blended binaries), and $6.18 \times 10^{9}$ fainter stars which serve as unresolved background stars. A catalog of simulated \tess detections from one run of the Monte Carlo simulations by \cite{Sullivan:apj2015a} contains the periods and radii of the detected planets as well as the stellar radii and effective temperatures of the host stars. We use this catalog to examine the likely stellar parameters of stars around which \tess will detect planets and ascertain the characteristics and impact of nearby stellar companions. As the primary goal of \tess is to find small planets with a measurable radial velocity signal, and it is difficult to measure precise radial velocities for stars fainter than $V = 12$, we only examine stars 12th magnitude or brighter. Stellar properties of the bright dwarf stars simulated by \cite{Sullivan:apj2015a} are shown in Figure \ref{fig:Shist}. The stellar masses are interpolated from the effective temperatures and radii provided in \citeauthor{Sullivan:apj2015a} and the Modern Mean Dwarf Stellar Color and Effective Temperature Sequence Table\footnote{Version 2017.10.19: \url{http://www.pas.rochester.edu/~emamajek/EEM_dwarf_UBVIJHK_colors_Teff.txt}} based on \cite{Pecaut:apjs2013a}.

\subsubsection{Companion detections by contrast}\label{subsec:sull_contrast}

\begin{figure*}[t]
\plottwo{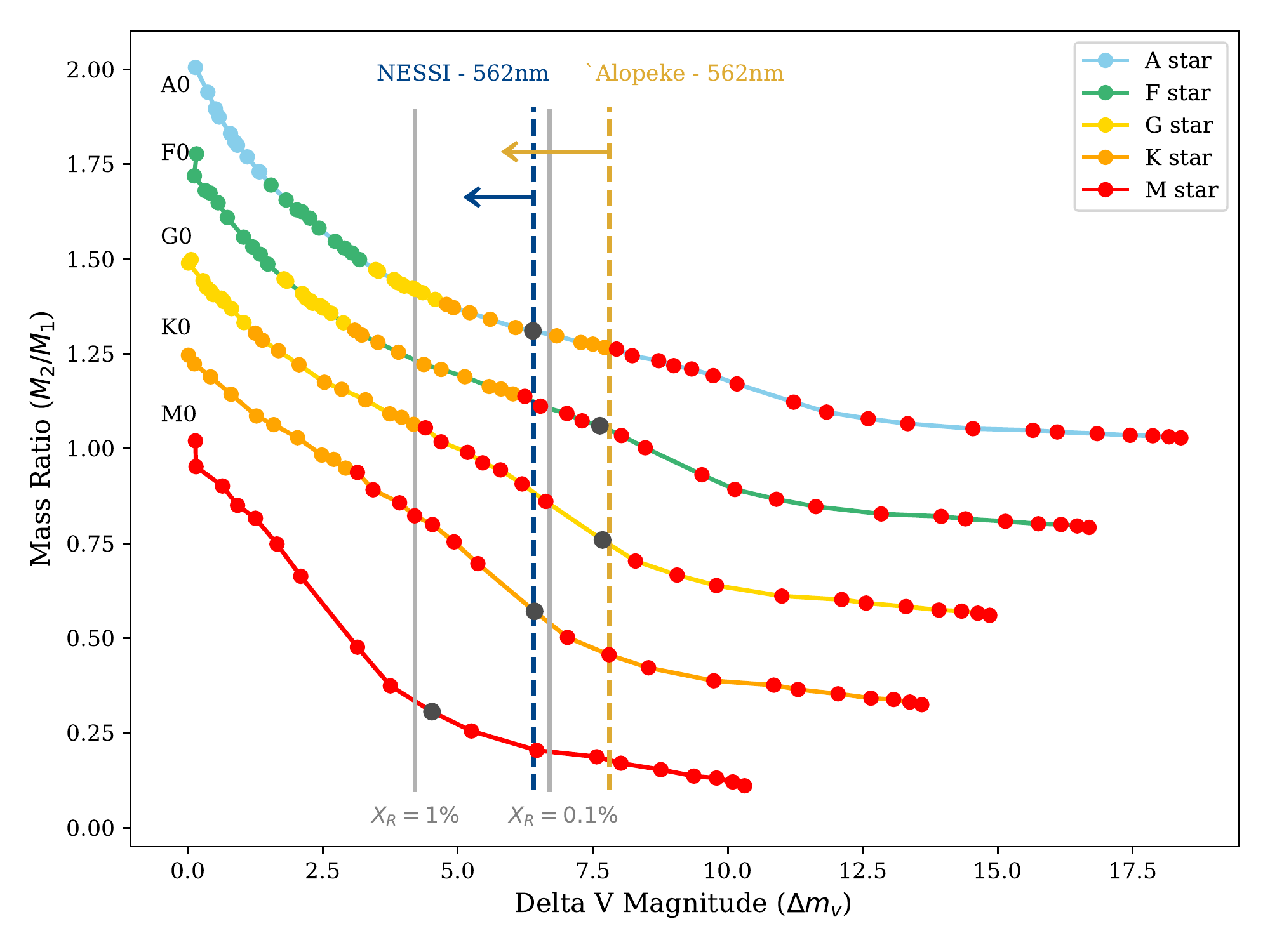}{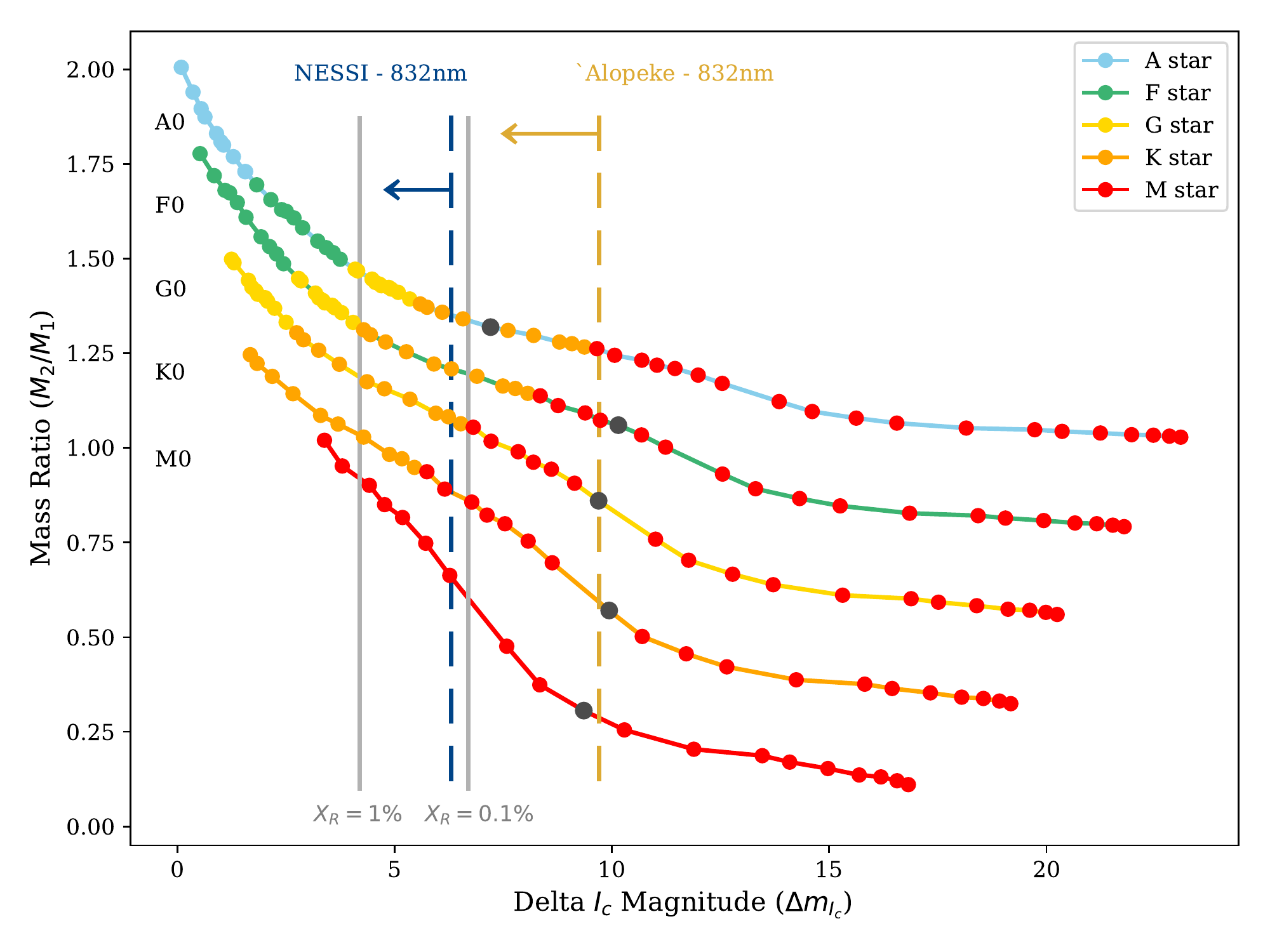}
\centering
\caption{Left: Mass ratio ($q = M_2/M_1$; offset by $q$ + 0.25 for clarity) vs.~$V$-band delta magnitude of possible companions for representative A0, F0, G0, K0, and M0 dwarfs from \cite{Sullivan:apj2015a}. The colors of the lines and dots correspond to the spectral type of the star/companion. Contrast limits for speckle imaging at 562nm with \nessi and \alopeke/\zorro are shown by the dashed blue line ($\Delta m \lesssim 6.4$) and dashed yellow line ($\Delta m \lesssim 7.8$), respectively. The solid gray lines show the contrast ratios ($\Delta m = 4.2, 6.7$) at which planet radii will have 1\% and 0.1\% correction factors ($X_R = 1.01, 1.001$) due to the presence of a stellar companion (see Section \ref{subsec:sull_contrast} for more details). The black dots highlight where the mass ratio for each star is $q \sim 0.3$. Right: Mass ratio (offset for clarity) vs.~$I_c$-band delta magnitude for the same stars and possible companions. Contrast limits for speckle imaging at 832nm with \nessi ($\Delta m \lesssim 6.3$) and \alopeke/\zorro ($\Delta m \lesssim 9.7$) are shown by the long dashed blue and yellow lines, respectively, while the solid gray lines again show the contrast ratios at which planet radii will have less than 1\% and 0.1\% correction factors due to a stellar companion. Speckle imaging can detect companion stars as faint as early M stars around A$-$G stars, and stars as faint as mid-M around K$-$M stars; however, the likelihood of detecting a companion also depends on the projected separation of the two stars (see Figures~\ref{fig:SulldistrG} \& \ref{fig:SulldistrW}). 
\label{fig:allcomps}}
\end{figure*}

To assess the types of companions speckle imaging can detect around \tess host stars, we consider all possible bound companions for the (simulated) stars in \cite{Sullivan:apj2015a} by selecting the nearest entry in terms of effective temperature in the Modern Mean Dwarf Stellar Color and Effective Temperature Sequence Table and assigning all stars with later spectral types (cooler effective temperatures) as possible companions. There is, in general, one entry in the table for each spectral type ($0-9$) for A, F, G, and K stars, and every 0.5 spectral type for M stars. We use the apparent $V$-band magnitude and distance modulus reported by \cite{Sullivan:apj2015a} to determine absolute visual magnitudes for each star, then subtract the absolute magnitude of each possible companion to get an estimated delta magnitude. We also determine delta magnitudes using apparent $I_{c}$ - band magnitudes, as the $I_{c}$ - band is roughly centered within the \tess badpass. The possible stellar companions for five representative stars from \cite{Sullivan:apj2015a} are illustrated in Figure~\ref{fig:allcomps}, where each line corresponds to a star of A0 (blue), F0 (green), G0 (yellow), K0 (orange), and M0 (red) spectral types and the filled circles represent possible companions. The mass ratio ($q = M_2/M_1$; offset by $q$ + 0.25 for clarity) is plotted as a function of magnitude difference from the primary ($\Delta m$) in $V$- (left) and $I_{c}$ - bands (right) for each possible companion. The colors of the lines and circles correspond to the spectral type of the star/companion. While the detection limits of speckle imaging depend on both the magnitude and separation of the companion star, the left hand plot highlights the magnitude difference obtainable at 562nm with \nessi ($\Delta m \lesssim 6.4$; dashed blue line) and \alopeke/\zorro ($\Delta m \lesssim 7.8$; dashed yellow line). The right hand plot shows the same stars and possible companions in the $I_c$ band, as well as the \nessi ($\Delta m \lesssim 6.3 $; long dashed blue line) and \alopeke/\zorro limits at 832nm ($\Delta m \lesssim 9.7$; long dashed yellow line). As shown in Figure~\ref{fig:allcomps}, speckle imaging can detect companion stars as faint as early M stars around A$-$F stars, and stars as faint as mid-M around G$-$M stars.

The binary mass ratio distribution determined by \cite{Raghavan:apjs2010a} for FGK stars is relatively flat (down to $q \sim 0.1$) for all mass ratios, with a $\sim$$2.5$x enhancement for nearly equal mass companion stars ($q > 0.95$). More specifically, \cite{Duchene:araa2013a} report that short-period binaries are characterized by a strong peak at $q \sim 1$ and a slowly declining function towards low mass ratios, while long-period binaries have a single peak around $q \sim 0.3$. Similarly, for M-type systems the mass ratio distribution is flat or slightly declining towards low-q systems, with short period M-dwarf binaries also biased towards high-q systems \citep{Duchene:araa2013a}. While only about $15\%$ of stars in the solar neighborhood have close, roughly equal mass companions \citep{Furlan:aj2017b}, due to their small $\Delta m$ we expect to detect nearly all such systems, as well as the peak in mass ratio for long-period systems (depending on the projected separation) since $q \sim 0.3$ (black dots) falls within the speckle contrast limits for all five spectral types shown in Figure~\ref{fig:allcomps}. 

Whether an unresolved binary is responsible for a transit event depends on the depth of the transit, as the transit depth provides a limit on the faintest blended binary that could produce a false positive signal matching the observed light curve. A fractional transit depth of $\delta = 0.01$ can be reproduced by blended systems 5 magnitudes fainter than the host star ($\Delta m$), whereas a transit of $\delta = 10^{-4}$ (approximately an Earth-sized transit of a solar-radius star) corresponds to a system with $\Delta m = 10$ \citep{Morton:apj2011a}. As seen in Figure~\ref{fig:allcomps}, speckle imaging will be able to detect companions with $\Delta m_{I_c} \lesssim 10$, potentially ruling out nearly all background eclipsing binaries that could mimic an Earth-sized transit for \tess planetary candidates. 

If a planet candidate is confirmed but one or more stellar companions are detected nearby, the extra flux from the companion(s) must be accounted for when deriving the planet radius. As the transit depth of a planet depends on the ratio of the squared radii  of the transiting object and the host star, the effect of the dilution can be accounted for and the radius corrected. \citet{Ciardi:apj2015a} defined a radius correction factor, $X_R$, as the ratio of the true planet radius to the observed planet radius. Assuming the planet orbits the primary/host star, a companion with $\Delta m = 1$ will result in the true planet radius being 1.18 times larger than the measured radius ($X_R = 1.18$), whereas a companion with $\Delta m = 5$ will only be overestimated by 1.005. Therefore, to get planet radii accurate to $\sim1\%$ ($X_R = 1.01$), any companions with a magnitude difference of $\Delta m \lesssim 4.3$ need to be detected and accounted for. The contrast ratios at which stellar companions will bias planetary radii by 1\% and 0.1\% ($X_R = 1.001$) are highlighted in Figure~\ref{fig:allcomps} by solid gray lines. Such companions are, in general, easily observable with speckle imaging, which will allow for the detection of nearly all stellar companions capable of diluting planetary transits. However, if there is more than one companion or the planet orbits a companion star, the radius correction factor can be $X_R = 2 - 5 $ for binary systems and even higher for triple systems, depending on the size and brightness of the companion star(s). Assuming planets are equally likely to orbit the primary or secondary stars, the mean correction factor for unvetted \kepler systems was determined to be $\sim1.5$ by \citet{Ciardi:apj2015a}, whereas the correction factor drops to $X_R = 1.2$ with high resolution vetting. The brighter, closer host stars of \tess will increase the effectiveness of such high-resolution imaging, however, decreasing the mean correction factor for vetted \tess candidates to $X_R \sim 1.07$ \citep{Ciardi:apj2015a}. \\

\subsubsection{Companion detections by separation}\label{subsec:sull_sep}

\begin{figure*}[t]
\includegraphics[scale=0.55]{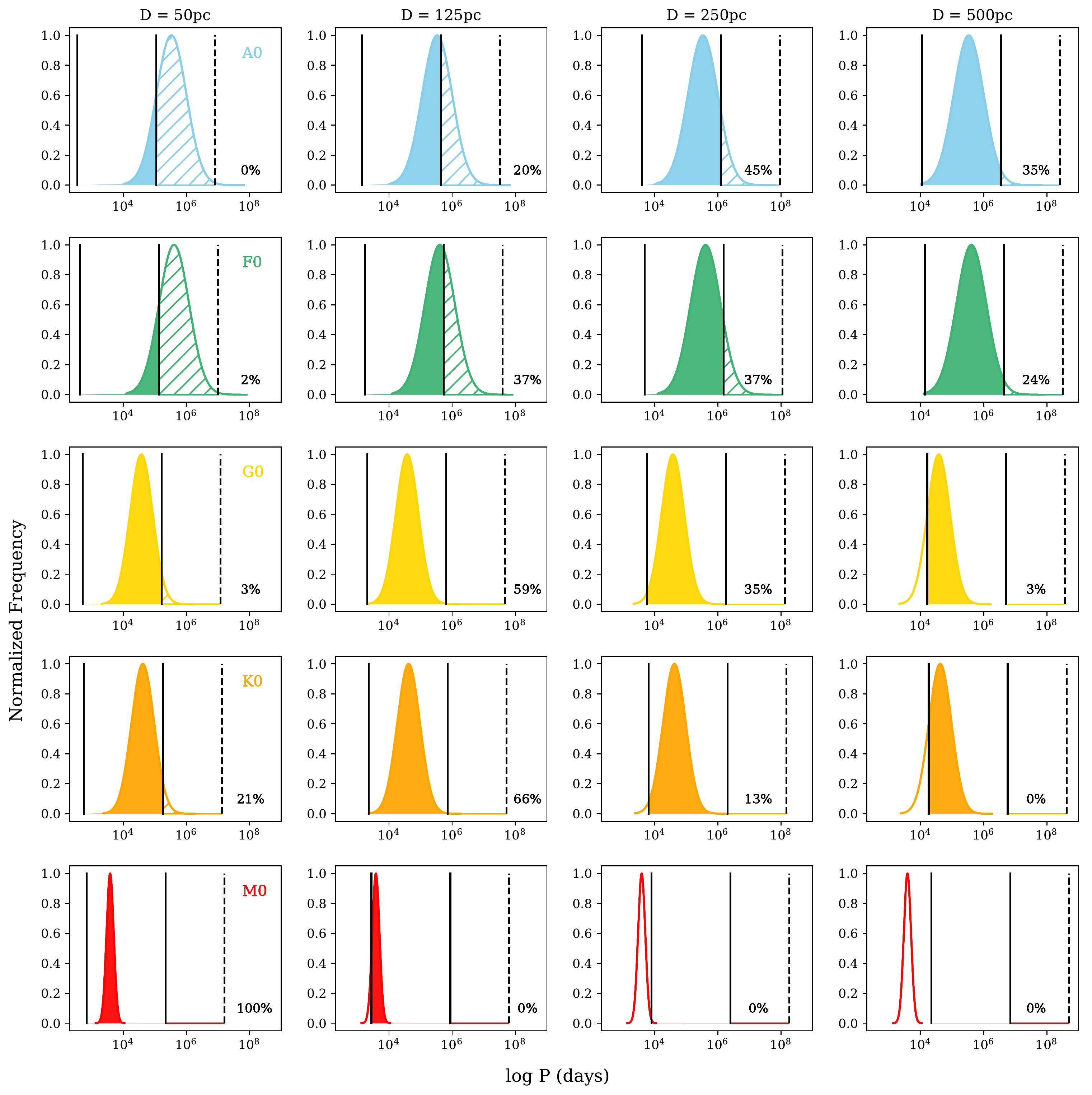}
\centering
\caption{Binary period distributions for representative A $-$ M (simulated) star primaries from \cite{Sullivan:apj2015a}. The distribution for M stars ($0.1 > M_\sun > 0.6$; red curve) peaks at close separations with a shallow tail that extends to large $\log P$. The shaded regions (color coded by spectral type of the primary) show the orbital periods corresponding to projected separations at which \alopeke/\zorro (832nm) can detect companions around stars at distances of 50, 125, 250, and 500pc. The solid black lines border the regions observable with speckle (0.026 - 1\farcs2), while the dashed lines highlight the \tess pixel size ($21''$) converted to $\log P$ space using the mass of the corresponding star. Companion stars beyond the speckle field of view of $1\farcs2$ will be detectable by other imaging techniques as indicated by the hatched regions. The fraction of stars for a given spectral type that fall within each distance range are noted as percentages on the individual subplots, e.g.~58\% of the G stars in \cite{Sullivan:apj2015a} have distances between 50 - 125pc, with six (unplotted) B stars located between 250 - 500 pc.
\label{fig:SulldistrG}}
\end{figure*}

\begin{figure*}[ht]
\includegraphics[scale=0.55]{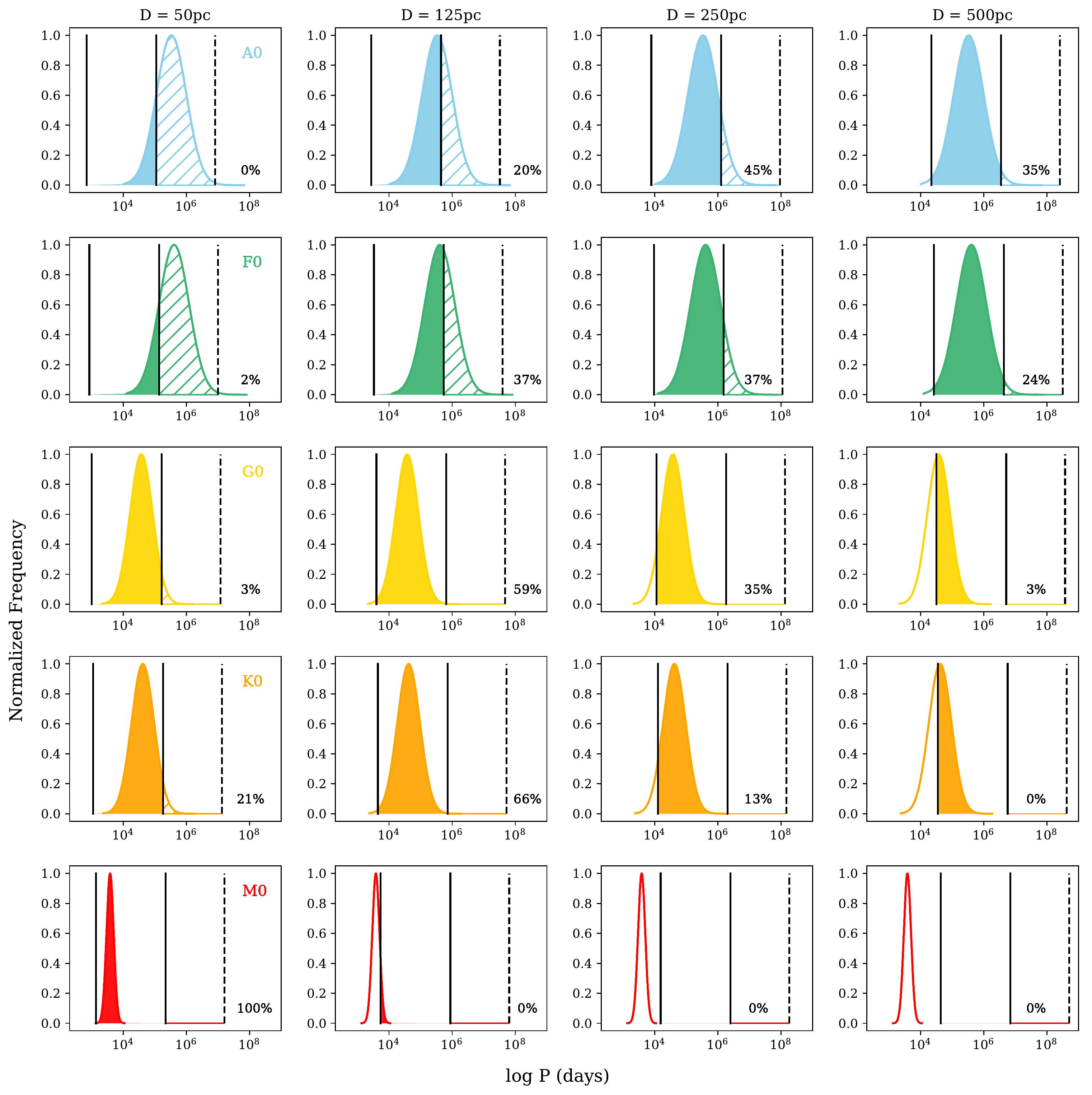}
\centering
\caption{Binary period distributions for representative A $-$ M star primaries from \cite{Sullivan:apj2015a}, similar to Figure~\ref{fig:SulldistrG}, depicting the orbital periods corresponding to the projected separations for which \nessi (562nm) can detect companions (0.04 - 1\farcs2). Due to the smaller aperture at WIYN, companion stars cannot be detected as close to the host star, resulting in more regions of parameter space not vetted by speckle imaging.
\label{fig:SulldistrW}}
\end{figure*}

When determining which stellar companions speckle imaging is sensitive to, the separation of the components must also be considered. Speckle imaging can detect companions at the diffraction limit of the telescope out to $\sim$$1\farcs2$, limited primarily by the need for correlated speckle patterns in the primary and secondary stars, which only occurs over small separations \citep[see][]{Horch:aj2012b, Horch:aj2017a}. At separations greater than $\sim1''$ other high-resolution imaging techniques, such as AO or wide-field speckle imaging \cite[see][]{Scott:pasp2018a, Casetti:2019a}, can be used to detect stellar companions. Since the projected separation of a stellar companion depends on the physical size of the companion's orbit, as well as the distance to the system, we use binary period distributions for A, F, G, K, and M stars to determine the likely orbits detectable with speckle imaging. For solar-type stars, the orbital period distribution is typically parameterized as a log-normal distribution with a maximum at $P \sim 250$ yr ($a \sim 45$ AU) and a dispersion of $\sigma_{\log P} \approx 2.3$ \citep{Duchene:araa2013a}. We use the mean semi-major axis (converted to mean period via Kepler's third law) and standard deviation of $\log P$ values for A$-$M stars as detailed in Table 3 of \cite{Sullivan:apj2015a} to determine the log-normal period distribution for each simulated star in \citeauthor{Sullivan:apj2015a} Binary period distributions, in terms of $\log P$, for the representative A0, F0, G0, K0, and M0 stars are plotted in Figure~\ref{fig:SulldistrG}. To convert the angular resolution of speckle imaging into $\log P$ limits (shown as solid black lines in Fig.~\ref{fig:SulldistrG}), we use distances of $D = 50, 125, 250,$ and 500 pc and the mass of each of the five representative stars (for Kepler's third law) from \cite{Sullivan:apj2015a}. The shaded region under each curve highlights the portion of $\log P$ space that is detectable using \alopeke/\zorro at 832 nm ($0.02 - 1\farcs2$), while the hatched areas depict the regions of $\log P$ space in which companions can be detected using other high-resolution and seeing-limited imaging techniques ($\sim$$1.2 - 21''$). Figure~\ref{fig:SulldistrW} shows the same representative A0, F0, G0, K0, and M0 stars at distances of $D = 50, 125, 250,$ and 500 pc, with the shaded regions depicting speckle limits of \nessi at 562 nm (0.04 - 1\farcs2) in $\log P$ space. The smaller aperture of WIYN, and therefore larger diffraction limit, means speckle imaging with \nessi is not as sensitive to companions around M-type stars at distances greater than $\sim$100 pc and short-period companions in G- and K-type stars at distances greater than $\sim200$ pc, respectively. 

Figure~\ref{fig:SulldistrG} demonstrates that for distances less than $\sim$125 pc, speckle imaging with \alopeke/\zorro is sensitive to a majority of companions around G-, K-, and M-type stars, but at distances greater than $\sim$150 pc the peak of the companion distribution for M stars cannot be angularly resolved. However,  \tess is not sensitive to small, short duration transits around such faint stars. On the other hand, speckle imaging will be able to detect the peak of the companion distribution for A $-$ K stars at distances greater than $\sim$$150$ pc. While not all companions with angular separations of $\sim$$0.02-1\farcs2$ will be detected due to contrast limits varying with separation, sub-optimal observing conditions, orbital phase effects, inclination effects, etc., examining the fraction of companion stars within the reach of speckle imaging is useful for assessing the parameter space probed with this technique and understanding the regions that can be vetted. 

Of the 682 stars in our sample from \cite{Sullivan:apj2015a}, 10\% have distances less than 50 pc, 48\% are between $50 - 125$pc, 30\% are between $125 - 250$pc, and 10\% have distances between $250 - 500$pc (see Figs.~\ref{fig:SulldistrG} \& \ref{fig:SulldistrW} for breakdown by spectral type and distance). Based on the companion orbital period distributions and the individual distances of the host stars in \citeauthor{Sullivan:apj2015a}, speckle imaging at 562nm using \alopeke/\zorro can detect  99\% of companions around 307 of the stars, 90\% of companions around 522 stars, and at least 50\% of companions around 621 stars. If we include the capabilities of other techniques to detect companions at separations greater than $1\farcs2$ (hatched regions in Figure~\ref{fig:SulldistrG}), only 48 (7\%) systems have companion distributions that are not fully vetted to $>90\%$. \\

\subsection{Barlcay et al.~2018} \label{subsec:barclay}

As the recent paper by \cite{Barclay:ArXiv2018a} revised the exoplanet yields expected from \tess using stars from the CTL, a prioritized subset of stars from the TESS Input Catalog \citep{Stassun:ArXiv2017a} deemed most suitable for the detection of small planets, we also examine the host star properties and expected stellar companions for their sample of predicted exoplanet host stars. The CTL contains $3.8 \times 10^{6}$ cool dwarf stars over the entire sky bright enough for \tess to observe, as well as scientifically valuable M dwarfs. \cite{Barclay:ArXiv2018a} determined which stars from version 6.1\footnote{The TIC and CTL are available from the MAST archive at \url{http://archive.stsci.edu/TESS}} of the CTL are observable with \tess and simulated likely 2-minute cadence targets based on their ranking in the CTL and a realistic distribution of targets among the different sectors \tess will observe. Then, they assigned zero or more planets to each of the $3.18 \times 10^{6}$ stars observable with \tess, resulting in predictions that \tess will detect $\sim$\,1300 transiting planets, 286 of which will be smaller than 2 Earth-radii and 46 smaller than 1.25 Earth-radii. 

\begin{figure*}[h]
\includegraphics[scale=0.6]{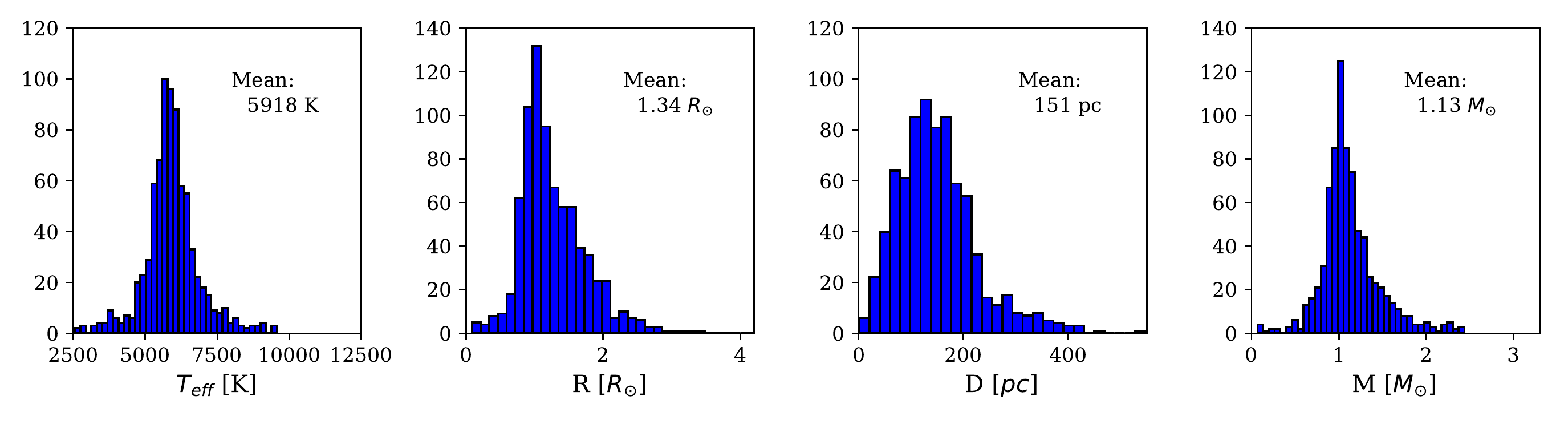}
\centering
\caption{Stellar properties of 787 bright dwarf stars ($V \leq 12$) with exoplanets detectable by \tess in simulated 2-minute cadence observations from \cite{Barclay:ArXiv2018a}. Stellar parameters are adopted from CTL v6.2, however there are 26 stars with no distance provided. The distributions are similar to those of \citet[][see Fig.~\ref{fig:Shist}]{Sullivan:apj2015a}, although the mean parameter values are slightly larger for stars in \citet{Barclay:ArXiv2018a}
\label{fig:Bhist}}
\end{figure*}

We use properties of the detected planets and host stars from the simulation that provided the median number of planets quoted in  \cite{Barclay:ArXiv2018a} to examine the likely stellar parameters of host stars and any possible companions, as well as to compare with the results of \cite{Sullivan:apj2015a}. We again trim the sample so that all stars are brighter than or equal to $V = 12$ and amenable to radial velocity measurements. Unlike the sample from \cite{Sullivan:apj2015a}, which contains only the planets detected in 2-minute cadence observations, the planet and host star properties given in \cite{Barclay:ArXiv2018a} contain stars with planets from the 2-minute cadence and full frame observations, resulting in nearly four times as many stars. As the 2-minute cadence results are more likely to produce targets for precision radial velocity and {\it{JWST}} follow-up, and to enable better comparison with \cite{Sullivan:apj2015a}, we exclude stars with planets detected via full frame observations. Stellar properties of the remaining 787 dwarf stars expected to produce transiting exoplanets are shown in Figure \ref{fig:Bhist}, and are similar to those of \citeauthor{Sullivan:apj2015a}~(see Fig.~\ref{fig:Shist}). \cite{Barclay:ArXiv2018a} note that roughly half of their target stars do not have distances reported in version 6.1 of the CTL, and a limited number have unrealistically large distances, but that these issues have been corrected in CTL v6.2. We therefore adopt the updated stellar parameters ($T_{\mathrm{eff}}$, $\log g$, radius, mass, and distance) from version 6.2 for our analysis; however, 26 of the target stars still have no reported distances.

\subsubsection{Companion detections}\label{subsec:barlcay_detect}

To assess which companions speckle imaging can detect around \tess stars from \cite{Barclay:ArXiv2018a}, we again consider all possible bound companions by selecting stars with later spectral types from the Modern Mean Dwarf Stellar Color and Effective Temperature Sequence Table. As the host stars used by \citeauthor{Barclay:ArXiv2018a} are real stars from the \tess Input Catalog, we reproduce the stellar properties from the CTL v6.2 for each star with  $V \leq12$ in Table~\ref{tab:comptable}, as well as provide limits on the companions speckle imaging can detect for each star. The first eight columns list the \tess Input Catalog ID, $M_*$, $R_*$, $T_{\mathrm{eff}}$, $V$- magnitude, \tess magnitude ($T$), distance, and spectral type inferred from the Modern Mean Dwarf Stellar Color and Effective Temperature Sequence Table for each star. The following two columns give the expected spectral types for companions with $\Delta m_v = 6$ and $8$ relative to the potential host star, typical \alopeke/\zorro speckle imaging contrast limits at 0\farcs1 and 1\farcs0, respectively. To estimate the fraction of possible companions that fall within the delta magnitude limits of speckle imaging for each star, we determine the number of companions with $\Delta m_v \leq 8$ from all companions that result in $q \geq 0.1$ and weight the likelihood of each companion by the mass ratio distribution of \citet{Raghavan:apjs2010a}, such that companions with mass ratios of $0.1 \leq q \leq 0.95$ are equally likely and those with $q > 0.95$ are enhanced by $2.5$x. This fraction of detectable companions for each star is listed in Table~\ref{tab:comptable} under `Comp.~Frac.' Since we subtract the absolute visual magnitude of each possible companion from the absolute visual magnitude of the corresponding star in \citet[][derived from apparent $V$ magnitude and distance in the CTL v6.2]{Barclay:ArXiv2018a}, there are slight differences in companion fractions between stars of the same spectral type. There are also three giant stars (TIC 27011422, TIC 359069654, TIC 445806156) whose properties are not well represented by The Modern Mean Dwarf Stellar Color and Effective Temperature Sequence, resulting in unrealistic companion spectral types and companion fractions.

Next, we examine the log-normal binary period distributions for the stars in \cite{Barclay:ArXiv2018a}. Converting the $\log P$ distribution to mean semi-major axis values via the mass and distance of each star, we use the limits of speckle imaging with \alopeke/\zorro at 562nm ($0.017 - 1\farcs2$) to determine the minimum and maximum physical separations at which companions may be detectable via speckle imaging; listed in Table~\ref{tab:comptable} as `Min Sep' and `Max Sep' in AU. Stars without distance estimates are listed for completeness, but no analysis of possible companions is included. Of the 787 stars from \cite{Barclay:ArXiv2018a} with $V \leq 12$, 6\% have distances less than  $50$ pc, 34\% have $50 <$ D $< 125$ pc, 48\% have $125 <$ D $< 250$pc, and 8\% have $250 <$ D $< 500$pc, with 2 stars at D $>$ 500 pc and 26 stars (3\%) with unknown distances. The fraction of the companion orbital period distribution observable with speckle imaging (similar to the shaded regions in Fig.~\ref{fig:SulldistrG}) is given in Table~\ref{tab:comptable} under `Distr.~Frac.'

\begin{deluxetable*}{lCCccccc|ccc|ccc|c}[t]
\tablecaption{Stellar parameters and companion space observable with speckle imaging for stars in Barclay et al.~2018 \label{tab:comptable}}
\tabletypesize{\footnotesize}
\tablecolumns{15}
\tablenum{2}
\tablewidth{0pt}
\tablehead{
\colhead{TIC-ID} & 
\colhead{$M_{*}$} & 
\colhead{$R_{*}$} & 
\colhead{$T_{\mathrm{eff}}$} &
\colhead{$V$} & 
\colhead{$T$} & 
\colhead{D} & 
\colhead{SpT} &
\multicolumn{2}{c}{Companion SpT at:} &
\colhead{Comp.} &
\colhead{Min Sep} & 
\colhead{Max Sep} &
\colhead{Distr.} &
\colhead{Speckle} 
\vspace{-2pt} \\
\cline{9-10}
 \colhead{} &
 \colhead{($M_{\sun}$)} &
 \colhead{($R_{\sun}$)} &  
 \colhead{$(K)$} &  
 \colhead{(mag)} & 
 \colhead{(mag)} & 
 \colhead{(pc)} & 
 \colhead{} &
 \colhead{$\Delta m_v = 6$} & 
 \colhead{$\Delta m_v = 8$} & 
 \colhead{Frac.} &
 \colhead{(AU)} & 
 \colhead{(AU)} & 
 \colhead{Frac.} & 
 \colhead{Frac.}
 }
\startdata
593228 & 0.253 & 0.268 & 3263 & 11.88 & 9.3 & 15.12 & M3.5V & M6V & M8V & 0.958 & 0.26 & 18.14 & 1.0 & 0.958 \\
682491 & 0.736 & 0.766 & 4702 & 8.13 & 7.05 & \nodata & K4V & \nodata & \nodata & \nodata & \nodata & \nodata & \nodata & \nodata \\
4026095 & 0.954 & 1.011 & 5458 & 10.62 & 9.78 & 133.39 & G8V & M2.5V & M4V & 0.865 & 2.33 & 160.07 & 0.989 & 0.856 \\
4064734 & 1.137 & 1.134 & 6076 & 10.29 & 9.77 & 148.69 & F9V & M2V & M3.5V & 0.97 & 2.6 & 178.43 & 0.993 & 0.963 \\
5645875 & 1.132 & 1.03 & 5649 & 8.76 & 8.14 & 51.12 & G5V & M3V & M4V & 0.889 & 0.89 & 61.34 & 0.712 & 0.632 \\
6029735 & 1.572 & 1.973 & 7090 & 6.45 & 6.2 & 64.27 & F1V & K7V & M2V & 0.91 & 1.12 & 77.12 & 0.019 & 0.018 \\
6077288 & 1.243 & 1.031 & 6314 & 8.87 & 8.34 & 71.84 & F6V & M2V & M3.5V & 1.0 & 1.25 & 86.21 & 0.879 & 0.879 \\
7218620 & 1.033 & 1.381 & 5480 & 10.2 & 9.63 & 137.42 & G8V & M2V & M3.5V & 0.8 & 2.4 & 164.9 & 0.99 & 0.792 \\
7320573 & 1.043 & 1.011 & 5804 & 11.11 & 10.58 & 176.3 & G2V & M2.5V & M4V & 0.968 & 3.08 & 211.55 & 0.997 & 0.965 \\
7444739 & 1.392 & 1.506 & 6688 & 9.33 & 8.9 & 158.3 & F3V & M0.5V & M3V & 1.0 & 2.76 & 189.95 & 0.995 & 0.995 \\
\enddata
\tablecomments{Table \ref{tab:comptable} is published in its entirety in the machine readable format.  A portion is shown here for guidance regarding its form and content.}
\vspace{-10pt}
\end{deluxetable*}

\begin{figure*}[h]
\plottwo{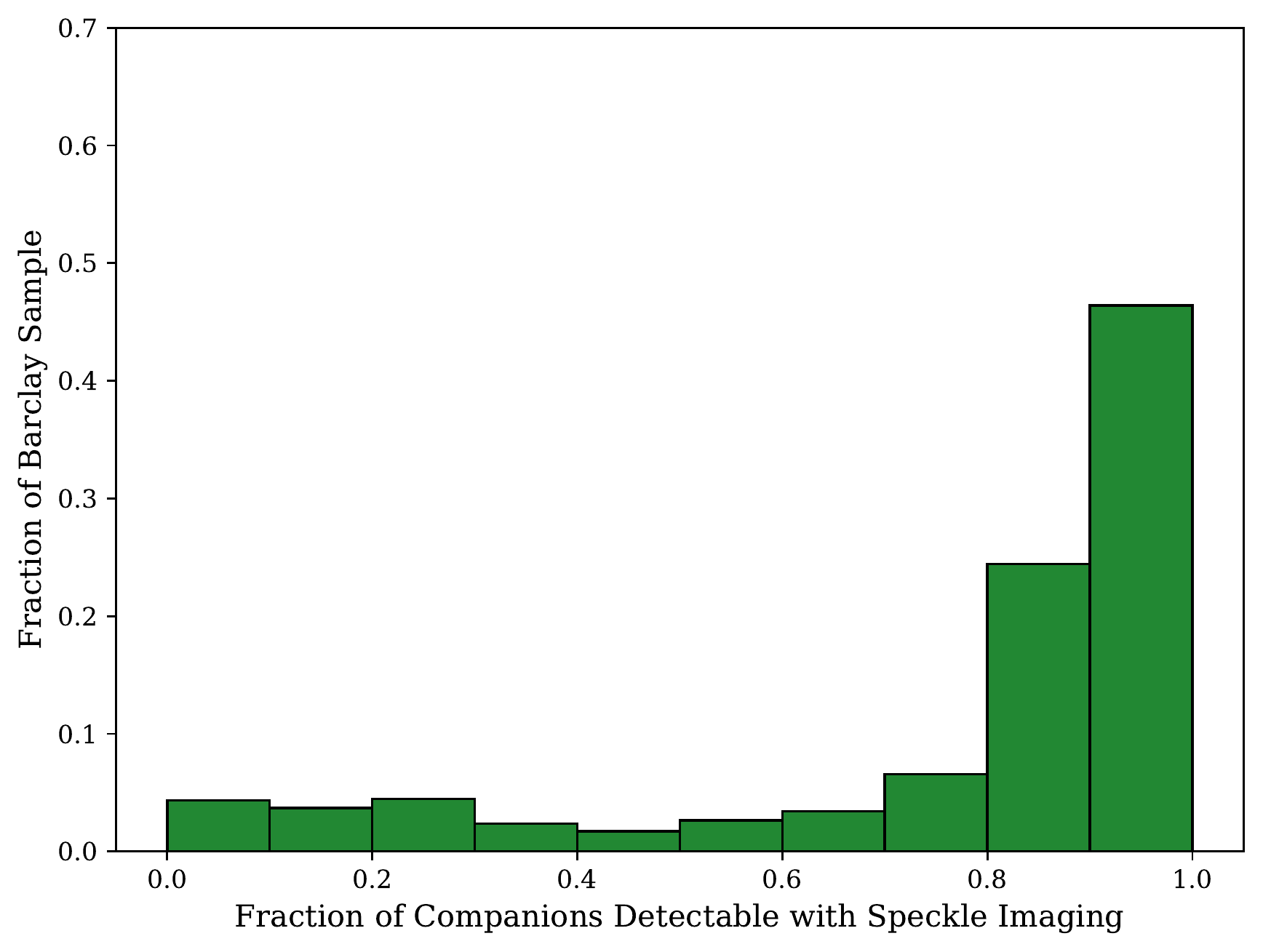}{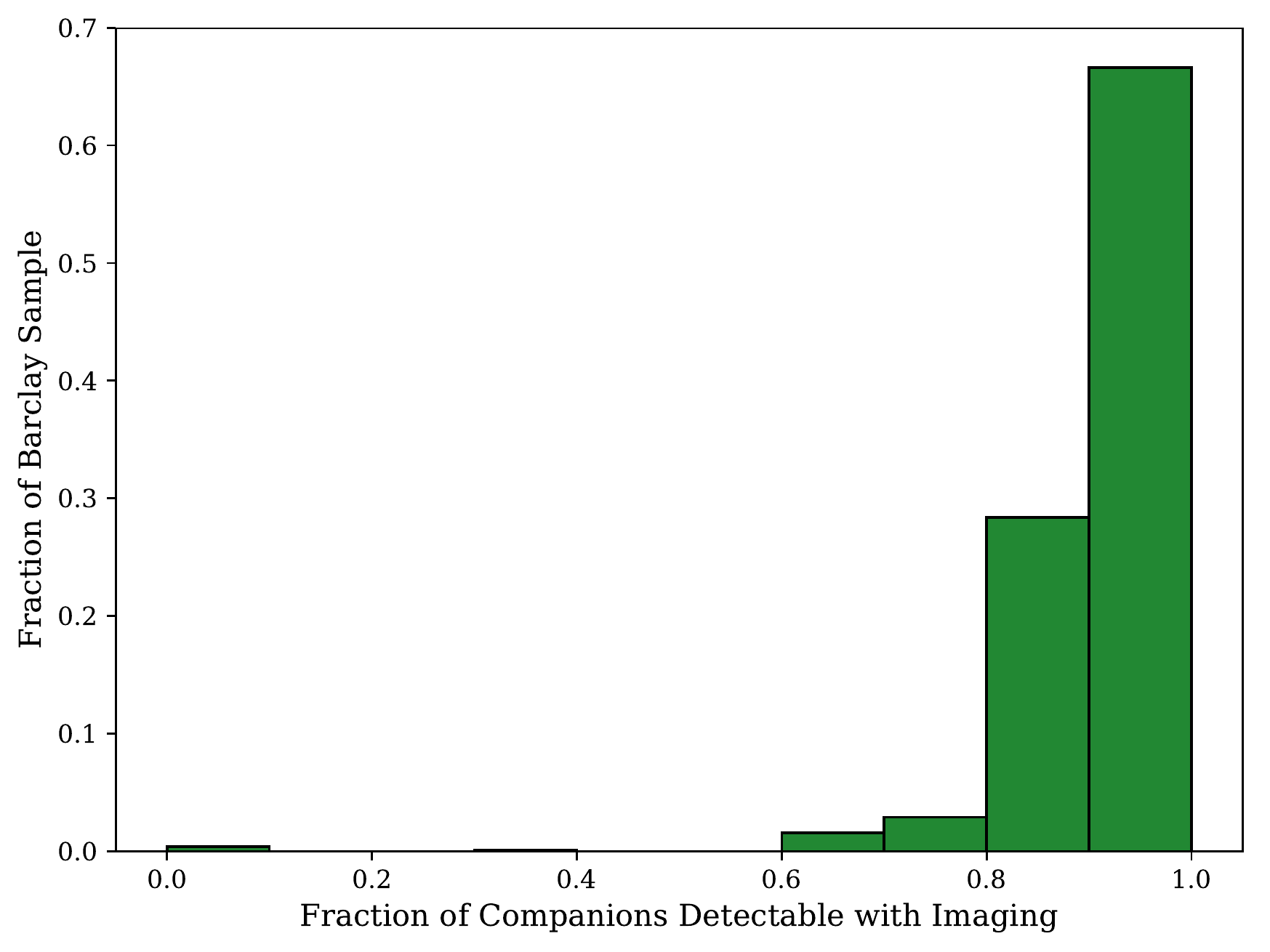}
\centering
\caption{Distribution of the estimated total fraction of possible companions that are detectable with speckle imaging around stars in \citet{Barclay:ArXiv2018a} with $V \leq 12$ (last column of Table~\ref{tab:comptable}). The histogram on the left is limited to the companions detectable within the $\sim0.02 - 1\farcs2$ angular resolution limits of speckle imaging (similar to the shaded regions in Figure~\ref{fig:SulldistrG}), while the right side shows the distribution of the fraction of companions detectable with speckle and other imaging techniques ($\sim 0.02 - 21''$; similar to the shaded + hatched regions of Figure~\ref{fig:SulldistrG}).
\label{fig:Balldet}}
\end{figure*}

\begin{figure*}[ht]
\includegraphics[scale=0.58]{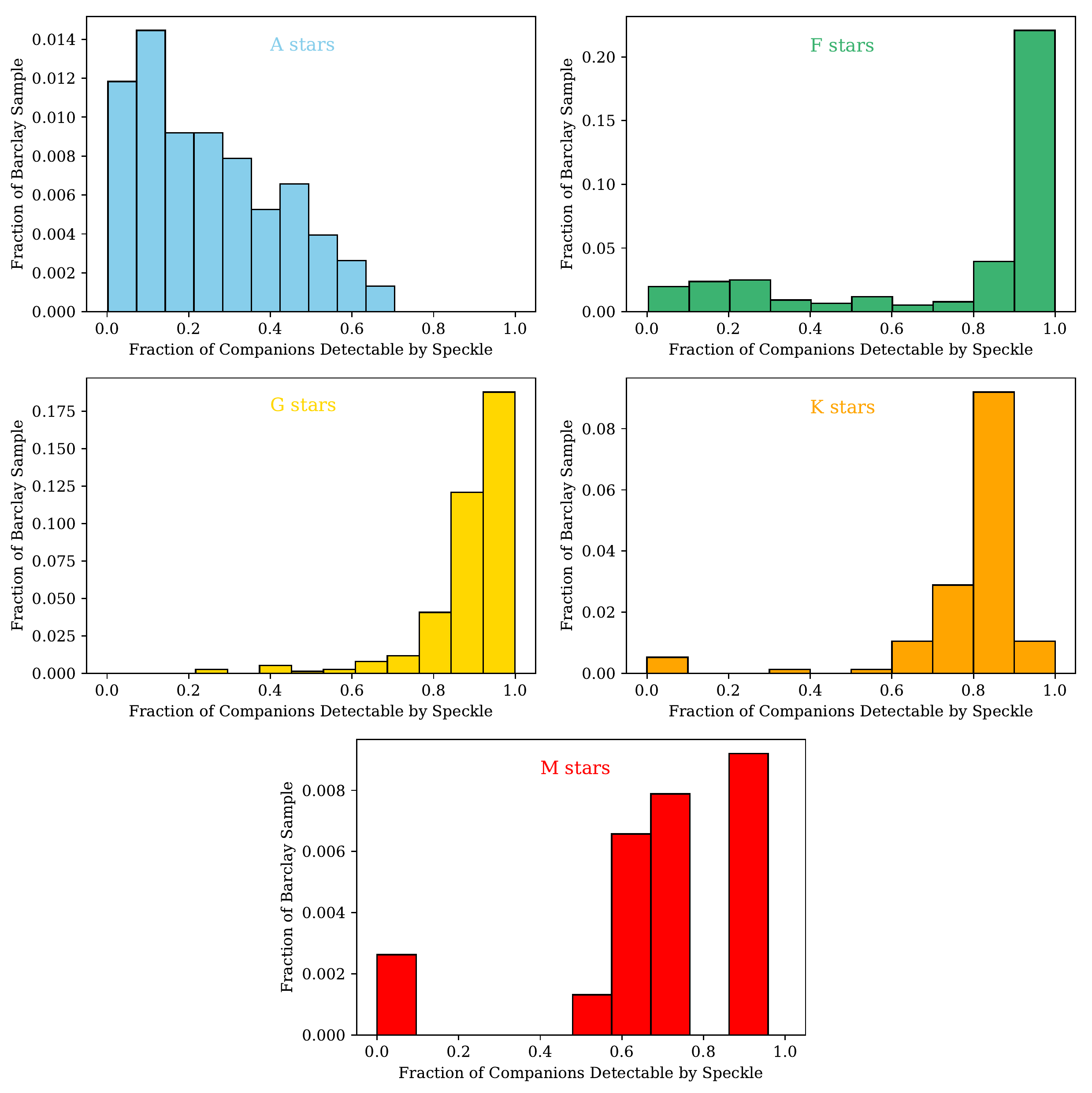}
\centering
\caption{Distributions of the fraction of possible companions detectable with speckle imaging around stars in \citet{Barclay:ArXiv2018a}, similar to the left hand side of Figure~\ref{fig:Balldet}, shown by spectral type.
\label{fig:Sptdet}}
\end{figure*}

To determine the total fraction of companion stars detectable with speckle imaging, we multiply the fraction of companions with $\Delta m$ less than the \alopeke speckle limit and the fraction of the companion period distribution observable with speckle (`Comp.~Frac.'~$\times$ `Distr.~Frac.'). The estimated fraction of detectable companions for each star in \citet{Barclay:ArXiv2018a} is listed in the last column of Table~\ref{tab:comptable} as `Speckle Frac'. Figure~\ref{fig:Balldet} shows the distribution of the total fraction of possible companions across all stars in the \citet{Barclay:ArXiv2018a} sample with $V \leq 12$. The histogram on the left is limited to the companions detectable within the $\sim0.02 - 1\farcs2$ angular resolution limits of speckle imaging (similar to the shaded regions in Figure~\ref{fig:SulldistrG}), while the right side shows the distribution of the fraction of companions detectable with speckle and other imaging techniques ($\sim 0.02 - 21''$; similar to the shaded and hatched regions of Figure~\ref{fig:SulldistrG}). The distribution of companions detectable with speckle imaging ($\sim0.02 - 1\farcs2$) for each spectral type is shown in Figure~\ref{fig:Sptdet}. For stars in \citet{Barclay:ArXiv2018a}, $99\%$ of all possible companions will be detected using \alopeke/\zorro at 562nm for 63 ($8\%$) of the 761 stars with known distances. For 353 ($46\%$) of the stars, $90\%$ of companions will be detected, and for 589 ($77\%$) stars at least 70\% of the companions will be detected with speckle imaging. If we include other techniques that are capable of detecting companions at separations greater than $1\farcs2$, $90\%$ of companions will be detected for 507 ($67\%$) of the \citet{Barclay:ArXiv2018a} stars, and $70\%$ of companions will be detected for 745 stars ($98\%$). 

\section{Discussion}\label{sec:disc}

The \tess planet yield simulations of \citet{Sullivan:apj2015a} and \citet{Barclay:ArXiv2018a} individually estimate that approximately 700 planets will be found around stars brighter than $V = 12$ via \tess 2-minute cadence observations. \citet{Sullivan:apj2015a} further estimate that more than 1000 astrophysical false positives are expected from the 2-minute cadence data. Assuming half of the astrophysical false positives have $V \leq 12$, roughly 1200 transiting events (excluding false positives due to instrumental effects) will need to be confirmed before more detailed characterization is possible. While the \tess 2-minute cadence targets are, in general, bright and nearby, it will take considerable time and resources to weed out false positives and identify candidates for further study. We have shown that speckle imaging is capable of detecting stellar companions within $\sim0.02-1\farcs2$ and $\Delta m \lesssim 7 - 9$ of planet candidate host stars, thereby eliminating some or all false positive scenarios due to background eclipsing binaries, as well as detecting excess flux that can dilute a planetary transit and lead to overestimated planetary radii. In addition to detecting nearby companions at relatively high contrasts, speckle imaging is extremely efficient, obtaining diffraction limited images in two colors for 30 $-$ 50 host stars per night. For the approximately 1200 planet candidates around bright stars that will need to be vetted before detailed RV and atmospheric follow-up is undertaken, speckle imaging will be able to observe them in roughly 30 nights. The speckle team typically devotes 4$-$8 nights per observing semester to exoplanet host star high-resolution imaging, and has been funded to continue to provide this community resource for the next three years, allowing for all of the bright \tess candidates to be observed within 2 $-$ 3 years.

As demonstrated in Section \ref{subsec:sull_contrast}, speckle imaging is capable of detecting stellar companions approaching $\Delta m = 10$ at 832nm, corresponding to the limiting magnitude for which \tess transits with fractional depths of $\sim$$10^{-4}$ can be caused by background binary stars. While speckle imaging will be unable to detect every companion around candidate host stars, the M-type stars that will typically be missed are unlikely to be the cause of such transit events. Faint companions that are close to the host star where the contrast limits for speckle are not as large may also remain undetected, although speckle will be able to provide tight constraints on such unresolved companions.Furthermore, the nearby stars targeted by \tess will increase the on-sky angular separation of stellar companions, enabling speckle imaging to probe physical separations of a few AU (see Table~\ref{tab:comptable}), the effective outer limit for spectroscopic sensitivity to binaries. Thus, with the addition of spectroscopy, essentially no stellar companion will remain undetectable. For stars that are bright enough (magnitudes $\lesssim$ 8), long-baseline interferometry also provides a way to detect companions at separations down to sub-milliarcsecond separations.
 
Even planet candidates that are able to be confirmed via other methods require high-resolution imaging to ensure there are no unresolved stars diluting the transit signal, as $\gtrsim 40$\% of exoplanet host stars have a bound companion \citep{Horch:apj2014a, Matson:aj2018a, Ziegler:aj2018b}. Without high-resolution imaging more than 95\% of the bound stellar companions ($< 1''$) remain unknown, and as a result, the derived planetary radii are systematically biased to smaller radii by a factor of $\sim1.1$x for \tess \citep{Ciardi:apj2015a}. Vetting with speckle imaging can detect and characterize bound stellar companions and, thus, reduce the radius bias. To ensure that any unresolved companions will not bias a planet's radius by more than 1\%, all companions with $\Delta m \lesssim 4.3$ need to be detected or ruled out. Such contrast ratios are obtainable with speckle imaging, allowing the light from close companions to be accurately deblended when calculating the true radius and density of identified planets, resulting in accurate planetary properties and occurrence rates.

In addition to spectroscopic observations, speckle imaging will be complemented by other high-resolution and seeing-limited imaging. The $21''$ pixels of \tess will require observations at a variety of angular scales to distinguish between multiple stellar sources contained within the \tess aperture as the source of any dips in brightness. {\it{Gaia}}, in particular, has already provided astrometry and photometry of more than a billion stars in the galaxy \citep{Gaia-Collab:aap2018a}, which will aid in the identification of nearby stars around \tess planet candidate host stars. {\it{Gaia's}} imaging resolution is $0.2''\times0.7''$, however, and not sensitive to low-contrast, sub-arcsecond companions, recovering only $\sim20$\% of companions detected by Robo-AO within $1''$ of \kepler planet hosts \citep{Ziegler:ArXiv2018a}. In general, Robo-AO is sensitive to companions with contrasts of $\Delta m < 4$ at separations of $0.5 - 1\farcs5$ and contrasts of $\Delta m < 7$ at $1.5 - 4\farcs0$ \citep{Ziegler:aj2018a}, and therefore not as sensitive to close-in and/or faint companions as speckle imaging. The use of multiple high-resolution imaging techniques, however, provides different sensitivities and resolutions which are valuable in ruling out as many false positives scenarios as possible and obtaining the most accurate orbital and physical parameters of the transiting object.

In addition to vetting planet candidates and yielding more accurate planetary radii, speckle imaging provides an assessment of stellar multiplicity for planet host stars that is crucial to understanding the role and influence of stellar companions on planet formation and evolution. Obtaining simultaneous images in two colors results in  robust determinations of multiple star position angles, separations, and magnitude differences, important for setting strong detection limits, as well as providing color information that can be used to estimate stellar parameters. Stellar parameters of the individual stars can be used to establish whether the detected companion is consistent with a bound companion by comparing the measured color to that predicted for a model bound companion via isochrones \citep[e.g.][]{Everett:aj2015a, Hirsch:aj2017a}. If the system is found to be a bound binary, then the companion may have played a dynamical role in the evolution of the planetary system and can provide clues to the role of stellar companions in planet formation and evolution. For instance, confirming the properties of such bound systems can elucidate the impact stellar companions have on planet formation and resolve inconsistencies in the stellar multiplicity rates of exoplanet host stars. 

Continued monitoring of detected companions with speckle imaging can also confirm whether the companions are physically bound via common proper motions or relative astrometry, and potentially lead to high-precision orbits and stellar masses for short-period systems. Long-term orbit monitoring of exoplanet host stars can further shed light on whether certain binary configurations are more likely for planet formation, as orbital parameters such as eccentricity and mutual inclination can impact the role of stellar companions in planet formation and planetary system architectures \citep{Dupuy:apj2016a}. Determining whether stellar companions are in the same plane as the transiting planets can therefore aid binary star and exoplanet formation and evolution models.

\section{Summary}\label{sec:conc}

We examine the ability of speckle imaging to detect stellar companions that will be unresolved in \tess data and with most other high-resolution imaging techniques. Using the simulated host star populations of \citet{Sullivan:apj2015a} and \citet{Barclay:ArXiv2018a}, we calculated the contrast ratios and binary period distributions of all possible stellar companions for each host star based on the Modern Mean Dwarf Stellar Color and Effective Temperature Sequence of \citet{Pecaut:apjs2013a}. By detecting companions with contrasts of $\Delta m \lesssim 7 - 9$, speckle imaging can detect companion stars as faint as early M stars around A$-$F stars and stars as faint as mid-M around G$-$M stars. This places strong constraints on blended eclipsing binaries as the source of brightness variations in \tess light curves, which will facilitate the rapid validation of hundreds of planet candidates and pave the way for further characterization efforts. Speckle imaging also enables de-blending of flux contributions from stellar companions, which ensures planetary radii are free from contamination at the $\sim1\%$ level and improves the accuracy of derived planet occurrence rates. With speckle imaging able to detect companions at separations of $\sim0.02-1\farcs2$, at least 90\% of the binary period distribution can be searched for nearly half of \tess planet candidate host stars with $V \leq 12$, and more than 50\% of the binary period distribution will be imaged in $\sim83\%$ of the host stars.

For the estimated 1200 planet candidates around bright stars that will be discovered in \tess 2-min cadence observations, speckle imaging will complete high-resolution follow-up observations in $2-3$ years. Reconstructed images and contrast limit curves produced by the speckle team, which will be available in the ExoFOP archives, will detect nearby stars that are not resolved in the \tess Input Catalog or by seeing-limited photometry. Such high-resolution imaging is critical for both exoplanet validation and for accurate characterization of planetary systems. Nearly 50\% of all exoplanet host stars are in multiple star systems, and the existence of companion stars must be accounted for when determining planetary properties. For transiting planets, high-resolution imaging can determine the flux dilution due to any companion stars, and therefore the true radius of the planet, necessary for deriving the bulk composition and atmospheric properties of the planet. Additionally, high-resolution imaging samples up to 99\% of the expected binary star distribution for systems located within a few hundred parsecs, allowing for studies of the influence of multiple stars on planetary demographics, formation, and evolution. Not accounting for the effects of stellar multiplicity also statistically biases planets toward smaller radii and gives rise to systematic errors in planet occurrence rates and completeness corrections.

\acknowledgments

The authors would like to thank all of the excellent staff at the WIYN and Gemini telescopes for their help during our observing runs. We gratefully acknowledge the support for our speckle imaging program given by the NASA Exoplanet Exploration Program and the role of the \kepler Science Office in upgrading \dssi to the two-EMCCD mode. This research has made use of the NASA Exoplanet Archive and the Exoplanet Follow-up Observation Program website, which are operated by the California Institute of Technology, under contract with NASA under the Exoplanet Exploration Program. RM's research was supported by an appointment to the NASA Postdoctoral Program at the NASA Ames Research Center, administered by Universities Space Research Association under contract with NASA.

\vspace{5mm}

\facilities{WIYN, Gemini:Gillett, Gemini:South}

\software{Python, Numpy, Matplotlib, SciPy, AstroPy, Jupyter Notebook}

\bibliography{/Users/rmatson/GoogleDrive/NPP/TeX/SpeckleLIB}

\end{document}